\documentclass[aps,prc,preprint,groupedaddress]{revtex4}

\usepackage{graphicx}

\begin{document}


\title{General classification and analysis of neutron beta-decay experiments}


\author{V. Gudkov }
\email[gudkov@sc.edu]{}
\affiliation{Department of Physics and Astronomy, University of South Carolina,
Columbia, SC 29208 }
\author{ G. L. Greene  }
\email[greenegl@ornl.gov]{}
\affiliation{Department of Physics,
University of Tennessee,
Knoxville, TN, 37996 }
\affiliation{Physics Division, Oak Ridge National Laboratory, Oak Ridge, TN, 37831}
\author{J. R. Calarco }
\email[calarco@unh.edu]{}
\affiliation{Department of Physics,
University of New Hampshire,
Durham, NH 03824}


\date{\today}

\begin{abstract}
A general analysis of the sensitivities of neutron
beta-decay experiments to manifestations of possible interaction beyond
the Standard Model is carried out.
 In a consistent fashion, we take
into account all known radiative and recoil corrections arising in the
 Standard Model. This provides a description of angular
correlations in neutron decay in terms of one parameter, which is accurate to the level of $\sim 10^{-5}$.  Based on this general
expression,  we present an analysis of the sensitivities to new physics  for selected neutron
decay experiments.
We emphasize that the usual parametrization of experiments in terms of the tree level coefficients $a$, $A$ and $B$ is inadequate when the experimental sensitivities are at the same or higher level relative to the size of the corrections to the tree level description.
\end{abstract}


\maketitle

\section{Introduction}

The relative simplicity of the decay of the free neutron makes it an
attractive laboratory for the study of possible extensions to the
Standard Model. As is well known, measurements of the neutron
lifetime and neutron decay correlations can be used to determine the
weak vector coupling constant, which, in turn, can be combined
with information on strange particle decay  to test such notions as the
universality of the weak interaction or to search for (or put a limit on)
nonstandard couplings (see, for example, \cite{gtw2,holsttr,deutsch,abele,yeroz,sg,herc,marc02} and references therein).
 It is less widely appreciated
that precision measurements of the correlations in neutron decay
can, in principle, be used as a test of the standard model without
appeal to measurements in other systems. In particular, the detailed
shape of the decay spectra and the energy dependence of the decay
correlation are sensitive to non-standard couplings. The extraction
of such information in a consistent fashion requires a rather
delicate analysis, as the lowest order description of the correlation
coefficients (and their energy dependencies) must be modified by a
number of higher order corrections that are incorporated within the
Standard Model. These include such effects as weak magnetism and
radiative corrections. Recently  \cite{eftcor} effective field theory
has been used to incorporate all standard model effects in a consistent
fashion in terms of one parameter with an estimated theoretical accuracy on the order of
$10^{-5}$. Because this accuracy is well below that anticipated in the
next generation of neutron decay experiments (see, for example, papers in \cite{NISTw}),
this analysis provides a useful framework for the exploration of the
sensitivity of various experiments to new physics.

In this paper, we extend the description of neutron beta-decay of \cite{eftcor} by including the most general  non-standard beta-decay interactions. Our framework provides a consistent description of the modifications of the beta-decay observables at a level well below
that anticipated in the next generation of experiments. Not
surprisingly, we find that the different experimental observables
have quite different sensitivities to the form of hypothetical
non-standard couplings (i.e. vector, scalar, etc.).

\section{Neutron $\beta$-decay beyond the Standard model. }

The most general description of neutron $\beta$-decay can be done in terms of low energy constants  $C_i$ by the Hamiltonian\cite{ly56,gtw1}
 \begin{eqnarray}
 H_{int}&=&(\hat{\psi}_p\psi_n)(C_S\hat{\psi}_e\psi_{\nu}+C^\prime_S\hat{\psi}_e\gamma_5\psi_{\nu})\nonumber \\
&+&(\hat{\psi}_p\gamma_{\mu}\psi_n)(C_V\hat{\psi}_e\gamma_{\mu}\psi_{\nu}+C^\prime_V\hat{\psi}_e\gamma_{\mu}\gamma_5\psi_{\nu})\nonumber \\
&+&\frac{1}{2}(\hat{\psi}_p\sigma_{\lambda\mu}\psi_n)(C_T\hat{\psi}_e\sigma_{\lambda\mu}\psi_{\nu}+C^\prime_T\hat{\psi}_e\sigma_{\lambda\mu}\gamma_5\psi_{\nu})\nonumber \\
&-&(\hat{\psi}_p\gamma_{\mu}\gamma_5\psi_n)(C_A\hat{\psi}_e\gamma_{\mu}\gamma_5\psi_{\nu}+C^\prime_A\hat{\psi}_e\gamma_{\mu}\psi_{\nu})\nonumber \\
&+&(\hat{\psi}_p\gamma_5\psi_n)(C_P\hat{\psi}_e\gamma_5\psi_{\nu}+C^\prime_P\hat{\psi}_e\psi_{\nu})  \label{ham} \\
&+& \text{Hermitian conjugate}, \nonumber
\end{eqnarray}
where the index $i=V$, $A$, $S$, $T$ and $P$ corresponds to vector,
axial-vector, scalar, tensor and pseudoscalar nucleon interactions.
In this presentation, the constants $C_i$ can be considered as  effective
constants of nucleon interactions with defined Lorentz structure,
assuming that all high energy degrees of freedom (for the Standard
model and any given extension of the Standard model) are integrated
out. In this paper we consider only time reversal conserving
interactions, therefore the constants $C_i$ can be chosen to be
real. (The violation of time reversal invariance in neutron decay at the level of considered accuracy would be a clear manifestation of new physics and thus does not require an analysis of the form contained here.) Ignoring electron and proton polarizations, the given
effective Hamiltonian will result in the neutron $\beta$-decay rate
\cite{gtw1}
 in the tree approximation (neglecting recoil corrections and radiative corrections)
in terms of the angular correlations coefficients $a$,  $A$, and $B$:
 \begin{eqnarray}
\frac{d\Gamma ^3}{dE_ed\Omega_ed\Omega_{\nu}}= \Phi (E_e)G_F^2
|V_{ud}|^2 (1+3\lambda^2)
\hskip 2cm \nonumber \\
\times (1+b\frac{m_e}{E_e}+a\frac{\vec{p}_e\cdot
\vec{p}_{\nu}}{E_e
E_{\nu}}+A\frac{\vec{\sigma} \cdot \vec{p}_e}{E_e}
+B\frac{\vec{\sigma} \cdot \vec{p}_{\nu}}{E_{\nu}}),
\label{cor}
\end{eqnarray}
Here,  $\vec{\sigma}$ is the neutron spin;
$m_e$ is the electron mass, $E_e$,
$E_{\nu}$, $\vec{p}_e$, and $\vec{p}_{\nu}$
are the energies and momenta of the electron and antineutrino,
respectively; and
$G_F$ is the Fermi constant of the weak interaction
(obtained from the $\mu$-decay rate).
The function $\Phi (E_e)$ includes normalization
constants, phase-space factors, and standard Coulomb corrections.
For the Standard model the angular
coefficients depend only on one parameter $\lambda = -C_A/C_V >0$,  the ratio of axial-vector to vector nucleon coupling constant ($C_V=C^\prime_V$ and $C_A=C^\prime_A$):
\begin{equation}
a=\frac{1-\lambda ^2}{1+3\lambda ^2},  \hskip 1cm A= -2\frac{\lambda
^2-{\lambda}}{1+3\lambda ^2}, \hskip 1cm B= 2\frac{\lambda
^2+{\lambda}}{1+3\lambda ^2}. \label{coef}
\end{equation}
 (The parameter $b$ is equal to zero for  vector - axial-vector  weak interactions.)

As  was shown in \cite{gtw2} the existence of additional
interactions  modifies the above expressions and can lead to a
non-zero value for the coefficient $b$. To explicitly see the influence of a
non-standard interaction  on the angular coefficients and on the decay rate
of neutron one can re-write the coupling constants $C_i$ as a sum of
a contribution from the standard model $C^{SM}_i$ and a possible
contribution from new physics $\delta C_i$:
 \begin{eqnarray}
C_V &=& C^{SM}_V + \delta C_V \nonumber \\
C^\prime_V &=& C^{SM}_V + \delta C^\prime_V \nonumber \\
C_A &=& C^{SM}_A + \delta C_A \nonumber \\
C^\prime_A &=& C^{SM}_A + \delta C^\prime_A \nonumber \\
C_S &=&  \delta C_S \nonumber \\
C^\prime_S &=&  \delta C^\prime_S \nonumber \\
C_T &=&  \delta C_T \nonumber \\
C^\prime_T &=&  \delta C^\prime_T.
\label{consts}
\end{eqnarray}
We neglect the pseudoscalar coupling constants since we treat\cite{gtw1} nucleons nonrelativistically.
Defining the term proportional to the total decay rate in eq.(\ref{cor}) as $\xi = (1+3\lambda^2)$ one can obtain corrections to parameters $\xi$, $a$, $b$, $A$ and $B$ due to new physics as $\delta\xi$, $\delta a$, $\delta b$, $\delta A$ and $\delta B$, correspondingly. Then, using results of \cite{gtw2},
\begin{eqnarray}
\delta\xi &=&   {C^{SM}_V}(\delta C_V+\delta C^\prime_V )+ ({\delta C_V}^2+{\delta C^\prime_V}^2+{\delta C_S}^2+{\delta C^\prime_S}^2)/2 \nonumber \\
 &+& 3 [ {C^{SM}_A}(\delta C_A +\delta C^\prime_A)+ ({\delta C_A}^2+{\delta C^\prime_A}^2+{\delta C_T}^2+{\delta C^\prime_T}^2)/2], \nonumber \\
\xi \delta b  &=& \sqrt{1-\alpha^2}[{C^{SM}_V}(\delta C_S+\delta C^\prime_S )+\delta C_S \delta C_V+ \delta C^\prime_S \delta C^\prime_V \nonumber \\
&+& 3({C^{SM}_A}(\delta C_T +\delta C^\prime_T)+\delta C_T \delta C_A+ \delta C^\prime_T \delta C^\prime_A )], \nonumber \\
\xi \delta a &=&  {C^{SM}_V}(\delta C_V+\delta C^\prime_V )+({\delta C_V}^2+{\delta C^\prime_V}^2-{\delta C_S}^2-{\delta C^\prime_S}^2)/2 \nonumber \\
&-&{C^{SM}_A}(\delta C_A +\delta C^\prime_A)-({\delta C_A}^2+{\delta C^\prime_A}^2-{\delta C_T}^2-{\delta C^\prime_T}^2)/2, \nonumber \\
\xi \delta A &=& -2{C^{SM}_A}(\delta C_A+{\delta C^\prime_A}) + \delta C^\prime_A \delta C^\prime_A -\delta C^\prime_T \delta C^\prime_T   \nonumber \\
&-& [C^{SM}_V(\delta C_A +\delta C^\prime_A)+{C^{SM}_A}(\delta C_V+\delta C^\prime_V )+\delta C_V \delta C^\prime_A +\delta C^\prime_V \delta C_A-\delta C_S \delta C^\prime_T -\delta C^\prime_S \delta C_T], \nonumber \\
\xi \delta B &=&   \frac{m \sqrt{1-\alpha^2}}{E_e}[2{C^{SM}_A}(\delta C_T+\delta C^\prime_T)+{C^{SM}_A}(\delta C_S+\delta C^{\prime}_S) + {C^{SM}_V}(\delta C_T+C^\prime_T)  \nonumber \\
 &+& 2 \delta C_T \delta C^\prime_A +2 \delta C_A \delta C^\prime_T +\delta C_S \delta C^\prime_A +\delta C_A \delta C^\prime_S + \delta C_V \delta C^\prime_T +\delta C_T \delta C^\prime_V]        \nonumber \\
&+&2{C^{SM}_A}(\delta C_A+{\delta C^\prime_A})-C^{SM}_V(\delta C_A +\delta C^\prime_A)-{C^{SM}_A}(\delta C_V+\delta C^\prime_V )  \nonumber \\
&-& \delta C_S \delta C^\prime_T - \delta C_T \delta C^\prime_S - \delta C_V \delta C^\prime_A - \delta C_A \delta C^\prime_V.
\label{nphys}
\end{eqnarray}

It should be noted that we have neglected radiative corrections and recoil effects for the new physics contributions, because these  are expected to be very small.  However, Coulomb corrections for the new physics contributions are taken into account since they are important for a low energy part of the electron spectrum.

From the above equations one can see that contributions from possible
new physics to the neutron decay distribution function  is rather complicated.
 To be able to separate new physics from different corrections to the expression (\ref{cor}), obtained in the tree  level of approximation,  one must   describe the neutron decay
 process with accuracy which is  better than the expected experimental accuracy.
 Assuming that the accuracy in future neutron decay experiments can reach a level
  of about $10^{-3} - 10^{-4}$, we wish to describe the neutron decay with theoretical
  accuracy by about $10^{-5}$ and our description must include all recoil and radiative
  corrections \cite{bilenky,sirlin,holstein,sirlinnp,sirlinrmp,garcia,wilkinson,sir,marciano}. To do this we will use recent results of calculations \cite{eftcor} of
  radiative corrections for neutron decay in the effective field theory (EFT) with some
   necessary modifications. The results of \cite{eftcor} can be used since they take into
   account both recoil and radiative corrections in the same framework of the EFT
   with estimated theoretical accuracy which is better than $10^{-5}$.  However, the EFT
   approach does not provide all  parameters but rather gives a
   parametrization in terms of a few  (two, in the case of neutron decay) low energy constants
   which must be extracted from independent experiments. Therefore,  the neutron $\beta$-decay
    distribution function is parameterized in terms of one unknown parameter (the second parameter is effectively
    absorbed in the axial vector coupling constant). If this parameter would be extracted from an independent experiment, it gives a model independent description of neutron beta-decay in the standard model with accuracy better than $10^{-5}$.  A rough estimate of this parameter based on  a ``natural'' size of strong interaction contribution to radiative corrections gives an accuracy for the expressions for the rate and the angular correlation coefficients which is better than $10^{-3}$ (see \cite{eftcor}).  We vary the magnitude of this parameter in a wide range for the given numerical analysis and show that variations of the parameters in the allowed range do not significantly change our results at a level well bellow $10^{-3}$.  Also, unlike \cite{eftcor}, we use the exact Fermi function for numerical calculations to take into account all corrections due to interactions with the classical electromagnetic field.
This gives us the expression for neutron decay distribution function as
 \begin{eqnarray}
\lefteqn{
\frac{d\Gamma ^3}{dE_ed\Omega_{\hat{p}_e}d\Omega_{\hat{p}_\nu}}
=
\frac{(G_FV_{ud})^2}{(2\pi)^5}
|\vec{p}_e|E_e(E_e^{max}-E_e)^2 F(Z,E_e)
}
\nonumber \\ && \times \left\{
f_0(E_e)
+\frac{\vec{p}_e\cdot\vec{p}_\nu}{E_eE_\nu}f_1(E_e)
+\left[\left(\frac{\vec{p}_e\cdot\vec{p}_\nu}{E_eE_\nu}\right)^2
-\frac{\beta^2}{3}
\right]f_2(E_e)
\right.  \nonumber\\ && \left.
+ \frac{\vec{\sigma}\cdot\vec{p}_e}{E_e}f_3(E_e)
+ \frac{\vec{\sigma}\cdot\vec{p}_e}{E_e}
\frac{\vec{p}_e\cdot\vec{p}_\nu}{E_eE_\nu}f_4(E_e)
+ \frac{\vec{\sigma}\cdot\vec{p}_\nu}{E_\nu}f_5(E_e)
+ \frac{\vec{\sigma}\cdot\vec{p}_\nu}{E_\nu}
\frac{\vec{p}_e\cdot\vec{p}_\nu}{E_eE_\nu}f_6(E_e)
\right\},
\label{eq;theresult}
\end{eqnarray}
where the energy dependent
angular correlation coefficients are:
\begin{eqnarray}
\lefteqn{f_0(E_e) = (1+3\lambda^2) \left( 1
+ \frac{\alpha}{2\pi} \delta_\alpha^{(1)}
+ \frac{\alpha}{2\pi} \; e_V^R \right) }
\nonumber \\ &&
- \frac{2}{m_N}\left[
 \lambda(\mu_V+\lambda)\frac{m_e^2}{E_e}
+\lambda(\mu_V+\lambda)E_e^{max}
-(1+2\lambda\mu_V+5\lambda^2)E_e
\right] ,
\\
\lefteqn{f_1(E_e) = (1-\lambda^2)
\left( 1
+ \frac{\alpha}{2\pi}
(\delta_\alpha^{(1)}+\delta_\alpha^{(2)})
+ \frac{\alpha}{2\pi} \; e_V^R \right)  }
\nonumber \\ &&
+\frac{1}{m_N}\left[
2\lambda(\mu_V+\lambda)E_e^{max}
-4\lambda(\mu_V+3\lambda)E_e
\right],
\\
\lefteqn{f_2(E_e) =
-\frac{3}{m_N}(1-\lambda^2)E_e , }
\\
\lefteqn{f_3(E_e) = (-2\lambda^2+2\lambda) \left( 1
+\frac{\alpha}{2\pi} ( \delta_\alpha^{(1)}
+\delta_\alpha^{(2)} )
+ \frac{\alpha}{2\pi} \; e_V^R \right)  }
\nonumber \\ &&
+\frac{1}{m_N}\left[
(\mu_V+\lambda)(\lambda-1)E_e^{max}
+(-3\lambda\mu_V+\mu_V-5\lambda^2+7\lambda)E_e
\right],
\\
\lefteqn{f_4(E_e) =
\frac{1}{m_N}(\mu_V+5\lambda)(\lambda-1)E_e, }
\\
\lefteqn{f_5(E_e) = (2\lambda^2+2\lambda) \left(1
+\frac{\alpha}{2\pi} \delta_\alpha^{(1)}
+ \frac{\alpha}{2\pi} \; e_V^R \right)  }
\nonumber \\ &&
+\frac{1}{m_N}\left[
-(\mu_V+\lambda)(\lambda+1)\frac{m_e^2}{E_e}
-2\lambda(\mu_V+\lambda)E_e^{max}
\right. \nonumber \\ && \left.
+(3\mu_V\lambda+\mu_V+7\lambda^2+5\lambda)E_e
\right] ,
\\
\lefteqn{f_6(E_e) =
\frac{1}{m_N}\left[
(\mu_V+\lambda)(\lambda+1)E_e^{max}
-(\mu_V+7\lambda)(\lambda+1)E_e
\right] \;  .  }
\end{eqnarray}
Here $e_V^R$ is the finite renormalized
low energy constant (LEC) corresponding to the ``inner"
radiative corrections due to the
strong interactions in the standard QCD approach;
$F(Z,E_e) $ is the standard Fermi function;
and the functions
$\delta_\alpha^{(1)}$ and
$\delta_\alpha^{(2)}$ are:
\begin{eqnarray}
\delta_\alpha^{(1)} &=&
\frac12
+ \frac{1+\beta^2}{\beta} {\rm ln}\left(\frac{1+\beta}{1-\beta}\right)
- \frac{1}{\beta}{\rm ln}^2\left(\frac{1+\beta}{1-\beta}\right)
+ \frac4\beta L\left(\frac{2\beta}{1+\beta}\right)
\nonumber \\ &&
+ 4 \left[\frac{1}{2\beta}{\rm ln}\left(\frac{1+\beta}{1-\beta}\right)
-1\right]
\left[{\rm ln}\left(\frac{2(E_e^{max}-E_e)}{m_e}\right)
+ \frac13 \left(\frac{E_e^{max}-E_e}{E_e}\right)
-\frac32
\right]
\nonumber \\ &&
+ \left(\frac{E_e^{max}-E_e}{E_e}\right)^2 \frac{1}{12\beta}
{\rm ln}\left(\frac{1+\beta}{1-\beta}\right) \, .
\\
\delta_\alpha^{(2)} &=&
\frac{1-\beta^2}{\beta}{\rm ln}\left(\frac{1+\beta}{1-\beta}\right)
+\left(\frac{E_e^{max}-E_e}{E_e}\right)
\frac{4(1-\beta^2)}{3\beta^2}
\left[\frac{1}{2\beta}{\rm ln}\left(\frac{1+\beta}{1-\beta}\right)-1
\right]
\nonumber \\ &&
+\left(\frac{E_e^{max}-E_e}{E_e}\right)^2
\frac{1}{6\beta^2}
\left[\frac{1-\beta^2}{2\beta}
{\rm ln}\left(\frac{1+\beta}{1-\beta}\right)-1
\right] \; ,
\end{eqnarray}
where $\beta = p_e/E_e$.
The only unknown parameter $e_V^R$ is chosen to satisfy the estimate \cite{sir} for an ``inner'' part of the radiative corrections:  $\frac{\alpha}{2\pi} \; e_V^R=0.02$.
In Eq.(\ref{eq;theresult}) the custom of
expanding the nucleon recoil correction
of the three-body phase space has been used.
These recoil corrections are included
in the coefficients $f_i$, $i=0, 1, \cdots , 6$
defined in the partial decay rate expression,
Eq.(\ref{eq;theresult}).
It should be noted that the expression for
$f_2$ is an exclusive
three-body phase space recoil
correction, whereas all other $f_i$, $i= 0, 1, 3, \cdots , 6$
contain a mixture of regular recoil and phase space
$(1/m_N)$ corrections.

The above expression presents all contributions from the Standard model. Therefore, the difference between this theoretical description and an experimental result can  only be due to effects not accounted for  by the Standard model. From the eqs.(\ref{nphys}) we can see that the only contributions from new physics in neutron decay are:
\begin{eqnarray}
f_0(E_e) &\longrightarrow & f_0(E_e) + \delta\xi + \frac{m}{E_e}\delta b, \nonumber \\
f_1(E_e) &\longrightarrow & f_1(E_e) + \delta a , \nonumber \\
f_3(E_e) &\longrightarrow & f_3(E_e) +  \delta A , \nonumber \\
f_5(E_e) &\longrightarrow & f_5(E_e) +  \delta B ,
\label{cphys}
\end{eqnarray}

Since possible contributions from models beyond the Standard one are rather
complicated, we have to use numerical analysis for
calculations of experimental sensitivities to new physics.

\section{The analysis of the experimental sensitivity to new physics}

To calculate the sensitivity of an experiment with a total number of events $N$  to the parameter $q$ we use the standard technique of the minimum variance bound estimator  (see, for example \cite{kend,frod}).  The estimated uncertainties provided by this method correspond to one sigma limits for a normal distribution.  The statistical error (variance) $\sigma_q$ of parameter $q$ in
 the given experiment can be written as
\begin{equation}\label{sen1}
  \sigma_q = \frac{K}{\sqrt{N}},
\end{equation}
where
\begin{equation}\label{sen2}
    K^{-2} = \frac{\int w(\vec{x})\left(\frac{1}{w(\vec{x})}\frac{\partial w(\vec{x})}{\partial q} \right)^2d\vec{x}}{\int w(\vec{x})d\vec{x}}.
\end{equation}
Here $w(\vec{x})$ is a distribution function of measurable
parameters $\vec{x}$. We can calculate the sensitivity of the
experiment to a particular
  coefficient $C_i$ or to a function of these coefficients.
  The results for these integrated sensitivities for each type of
  interaction ($C_i$) and for the left-right model  are given in the
  table \ref{ctab} for the standard experiments measuring $a$, $A$ and $B$
  coefficients in neutron decay, assuming that all coefficients $C_i$ have the same value of $1\cdot 10^{-3}$. The numerical test shows that results for the coefficients $K$  can be linearly re-scaled for the parameters $C_i$ in the range from $10^{-2}$ to $10^{-4}$ with an accuracy of better than $10 \%$.
  We can see that different experiments have
   different sensitivities (discovery potentials) for the possible manifestations
   of new physics.

\begin{table}
  \centering
  \caption{Relative statistical error ($K$) of the standard experiments to different types of interactions from new physics ($C_i$ constants) provided that these constants have the same values of $1\cdot 10^{-3}$.}
  \label{ctab}
\begin{tabular}{|c|c|c|c|}
\hline
 Interactions & $a$ & $A$ & $B$ \\
\hline
$V$ & 5.26 & 3.60 & 6.95 \\
$A$ & 1.73 & 1.90 & 1.91 \\
$T$ & 2.59 & 7.25 & 1.50 \\
$S$ & 8.70 & 26.70 & 1.46 \\
$V+A$ & 2.01 & 1.58 & 3.86 \\
\hline
\end{tabular}
\end{table}

The given description of neutron $\beta$-decay experiments in terms of low energy constants related to the Lorentz structure of weak interactions is  general and complete.  All models beyond the Standard one (new physics) contribute to the $C_i$ values in different ways. Therefore, each model can be described by a function of the  $C_i$ parameters. To relate these $C$-coefficients explicitly to the possible models beyond the Standard one we can use the parametrization of reference \cite{herc}.  It should be noted that  the definitions of reference \cite{gtw1} used for the $C_i$ coefficients are the same as in  \cite{herc}, except for the opposite sign of $C^\prime_V$, $C^\prime_S$, $C^\prime_T$ and $C_A$.  Therefore, we can re-write the relations of the $\delta C_i$, which contain contributions to the $ C_i$  from new physics, in terms of the parameters $\bar{a}_{jl}$ and $\bar{A}_{jl}$ defined in the paper \cite{herc} as:
\begin{eqnarray}
\delta C_V &=& C^{SM}_V (\bar{a}_{LL}+\bar{a}_{LR}+\bar{a}_{RL}+\bar{a}_{RR}), \nonumber  \\
\delta C^\prime_V &=& -C^{SM}_V (-\bar{a}_{LL}-\bar{a}_{LR}+\bar{a}_{RL}+\bar{a}_{RR}), \nonumber  \\
\delta C_A &=& -C^{SM}_A (\bar{a}_{LL}-\bar{a}_{LR}-\bar{a}_{RL}+\bar{a}_{RR}),  \nonumber \\
\delta C^\prime_A &=& C^{SM}_A (-\bar{a}_{LL}+\bar{a}_{LR}-\bar{a}_{RL}+\bar{a}_{RR}) \nonumber  \\
\delta C_S &=& g_S (\bar{A}_{LL}+\bar{A}_{LR}+\bar{A}_{RL}+\bar{A}_{RR}),  \nonumber \\
\delta C^\prime_S &=-& g_S (-\bar{A}_{LL}-\bar{A}_{LR}+\bar{A}_{RL}+\bar{A}_{RR}),  \nonumber  \\
\delta C_T &=& 2 g_T (\bar{\alpha}_{LL}+\bar{\alpha}_{RR}), \nonumber  \\
\delta C^\prime_T &=& -2 g_T (-\bar{\alpha}_{LL}+\bar{\alpha}_{RR}).
\label{carel}
\end{eqnarray}

The parameters $\bar{a}_{jl}$, $\bar{\alpha}_{jl}$ and $\bar{A}_{jl}$  describe
contributions to the low energy Hamiltonian from current-current
interactions in terms of $j$-type of leptonic current and $i$-type
of quark current. For example, $\bar{a}_{LR}$ is the contribution to
the Hamiltonian from left-handed leptonic current and right-handed quark current
normalized by the size of the Standard Model (left--left current)
interactions.
 $g_S$ and $g_T$ are formfactors at zero-momentum transfer in the nucleon matrix element of scalar and tensor currents.  For more details, see the paper \cite{herc}. It should be
noted, that $\delta C_i + \delta C^\prime_i $ involve left-handed
neutrinos and $\delta C_i - \delta C^\prime_i $ is related to right-handed
neutrino contributions in corresponding lepton currents. The analysis of the three
experiments under consideration ($a$, $A$ and $B$ coefficient measurements) in terms of
sensitivities ($K^{-1}$) to $\bar{a}_{jl}$, $\bar{\alpha}_{jl}$ and $\bar{A}_{jl}$
parameters is presented in the table \ref{atab}.  For the sake of easy comparison the sensitivities in this table  are calculated under assumptions that all parameters ($\bar{a}_{jl}$, $\bar{\alpha}_{jl}$ and $\bar{A}_{jl}$) have exactly the same value, $1\cdot 10^{-3}$. The expected values of these parameters vary over a wide range from $0.07$ to $10^{-6}$ (see table \ref{nptab} and,  paper \cite{herc} for the comprehensive analysis). The numerical results for the coefficients $K$ in the table can be linearly re-scaled for the parameters $\bar{a}_{ij}$, $\bar{\alpha}_{jl}$ and $\bar{A}_{ij}$ in the range from $10^{-2}$ to $10^{-4}$ with an accuracy better than $10 \%$.
The relative statistical errors presented in the Table demonstrate discovery potentials of different experiments to new physics in terms of parameters $\bar{a}_{ij}$, $\bar{\alpha}_{jl}$ and $\bar{A}_{ij}$. It should be noted, that the parameter $\bar{a}_{LR}$ cannot provide sensitive information on new physics at the quark level, unless we obtain the axial-vector coupling constant $g_A$ from another experiment, since in correlations $\bar{a}_{LR}$ appears in a product with $g_A$ (see \cite{herc}).  For discussion of significance of each of these parameters to models beyond the standard one see \cite{herc}.

\begin{table}[h]
  \centering
  \caption{Relative statistical error ($K$) of the standard experiments to different
  types of interactions from new physics ($\bar{a}_{ij}$ constants) provided that these constants have the same values of $1\cdot 10^{-3}$.}
  \label{atab}
\begin{tabular}{|c|c|c|c|c|c|c|c|c|c|c|}
  \hline
  & $\bar{a}_{LL}$ & $\bar{a}_{LR}$  & $\bar{a}_{RL}$ &$\bar{a}_{RR}$& $\bar{A}_{LL}$ & $\bar{A}_{LR}$ & $\bar{A}_{RL}$ & $\bar{A}_{RR}$ & $\bar{\alpha}_{LL}$ & $\bar{\alpha}_{RR}$\\
\hline
a & 0.17 & 0.25 & 135 & 487 & 1.43 & 1.43 & 283 & 283 & 0.19 & 79 \\
A & 1.53 & 0.63 & 423 & 1026 & 13.1 & 13.1 & 860 & 860 & 1.82 & 223 \\
B & 0.58 & 1.21 & 89 & 347 & 0.72 & 0.72 & 958 & 958 & 0.37 & 59 \\
\hline\end{tabular}
\end{table}

  It should be noted the results in the tables \ref{ctab} and \ref{atab} are calculated with the estimated value of the parameter $( \alpha /(2 \pi) \; e_V^R=0.02$. Numerical tests show that a change of this parameter by a factor two leads to changes of results in the tables by about $1\%$.

\begin{table}
  \centering
  \caption{Possible manifestations of new physics}
  \label{nptab}
\begin{tabular}{|c|c|c|c|c|c|c|}
  \hline
Model & L-R & Exotic Fermion & Leptoquark & Contact interactions & SUSY & Higgs \\
\hline
$\bar{a}_{RL}$ & 0.067 & 0.042 &  &  &  &  \\
$\bar{a}_{RR}$ & 0.075 &  & 0.01 &   &   &   \\
$\bar{A}_{LL}+\bar{A}_{LR}$ &   &   &  & 0.01 & $7.5 \cdot 10^{-4}$ & $3\cdot 10^{-6}$ \\
$\bar{A}_{RR}+\bar{A}_{RL}$ &   &   &   & 0.1 &   &   \\
$-\bar{A}_{LL}+\bar{A}_{LR}$ &   &   & $3\cdot 10^{-6}$ &   &   &  \\
$\bar{A}_{RR}-\bar{A}_{RL}$ &  &   & $4\cdot 10^{-4}$ &  &  &  \\
\hline\end{tabular}
\end{table}

\begin{figure}
\includegraphics{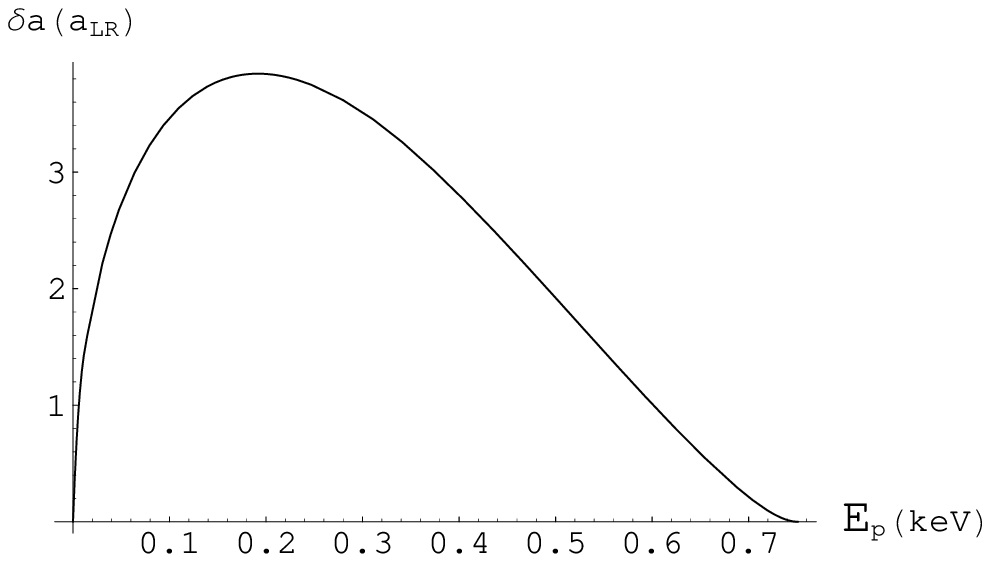}
\caption{Manifestation of $a_{LR}$-type interactions on the $a$ coefficient. }
\label{fig-a-aLR}
\end{figure}

\begin{figure}
\includegraphics{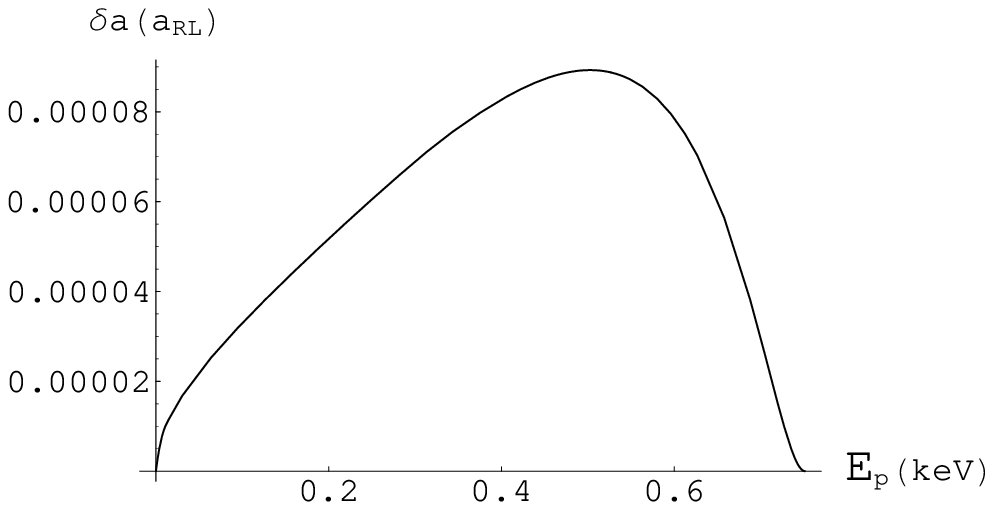}
\caption{Manifestation of $a_{RL}$-type interactions on the $a$ coefficient. }
\label{fig-a-aRL}
\end{figure}

\begin{figure}
\includegraphics{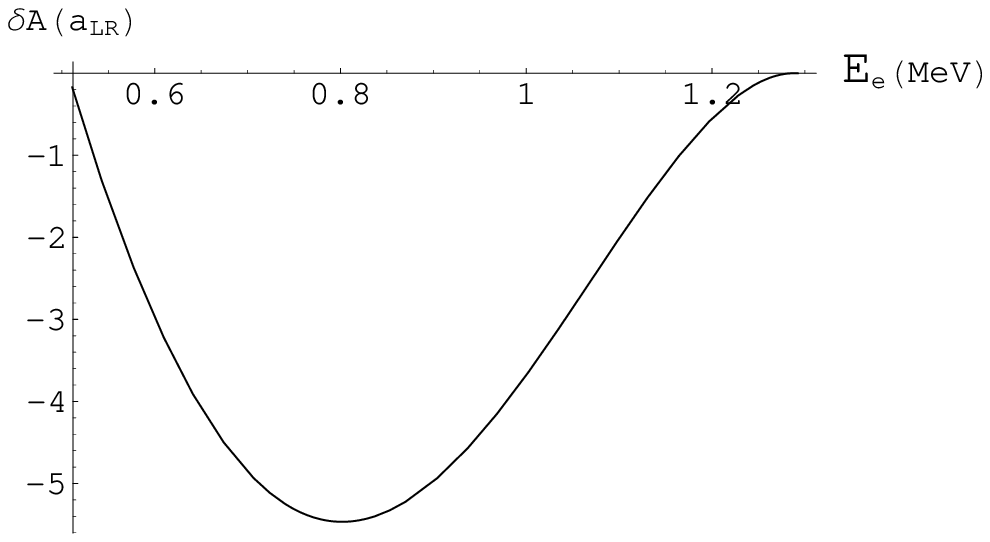}
\caption{Manifestation of $a_{LR}$-type interactions on the  $A$ coefficient. }
\label{fig-A-aLR}
\end{figure}

\begin{figure}
\includegraphics{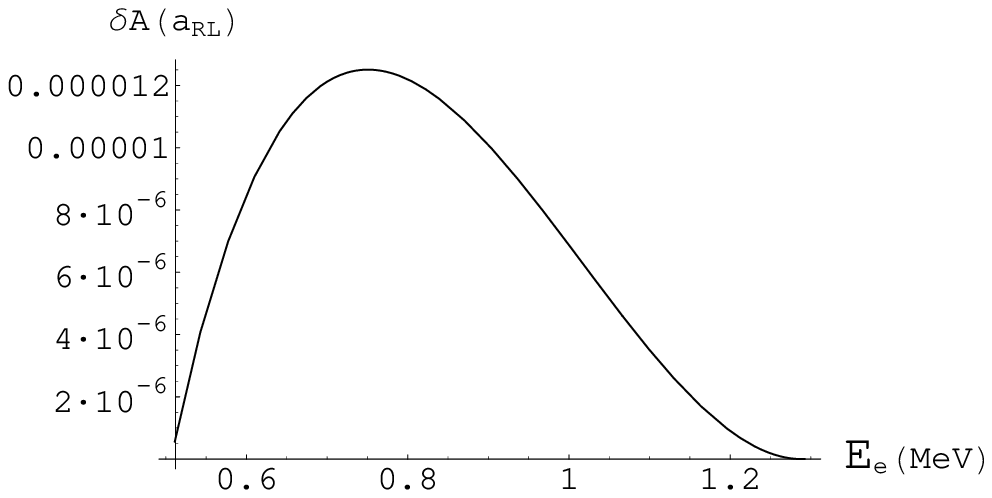}
\caption{Manifestation of $a_{RL}$-type interactions on the $A$ coefficient. }
\label{fig-A-aRL}
\end{figure}

\begin{figure}
\includegraphics{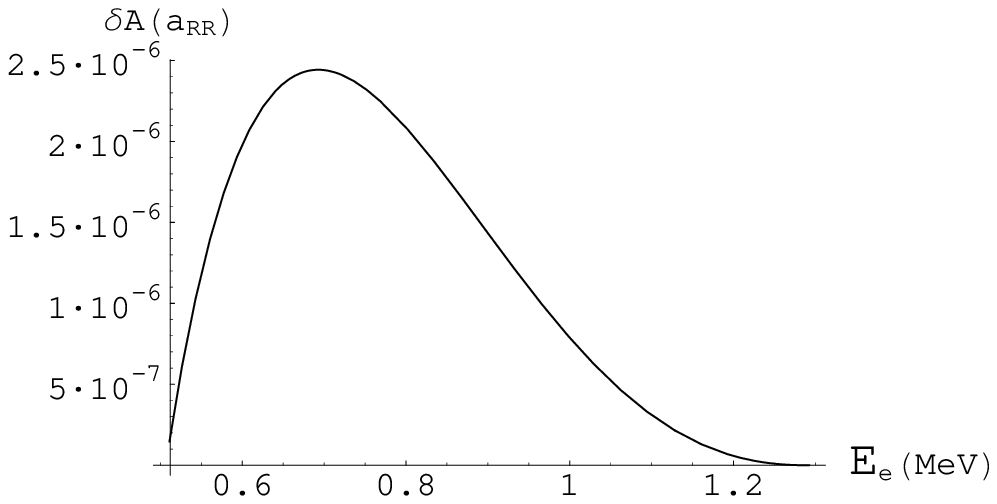}
\caption{Manifestation of $a_{RR}$-type interactions on the $A$ coefficient. }
\label{fig-A-aRR}
\end{figure}

\begin{figure}
\includegraphics{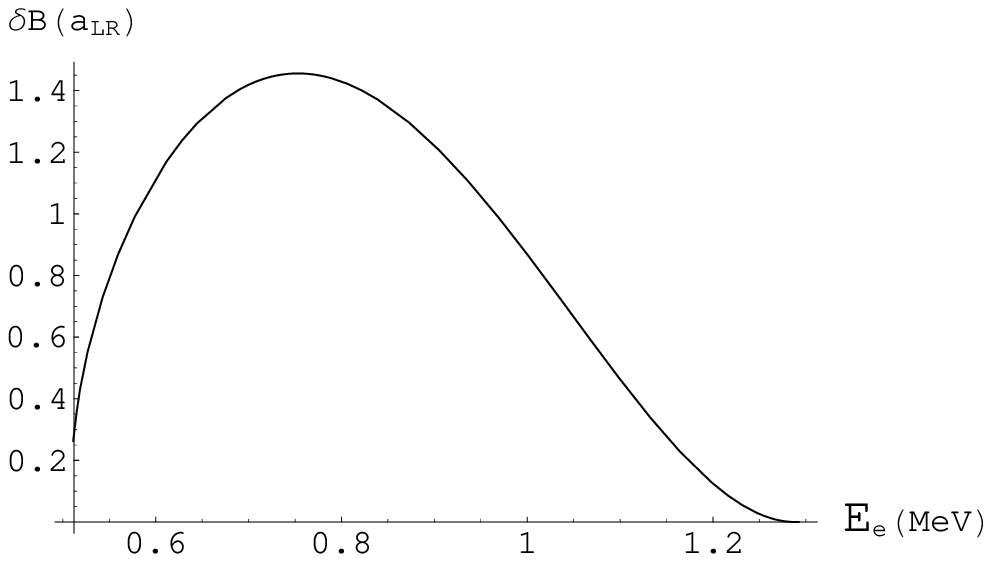}
\caption{Manifestation of $a_{LR}$-type interactions on the $B$ coefficient.  }
\label{fig-B-aLR}
\end{figure}

\begin{figure}
\includegraphics{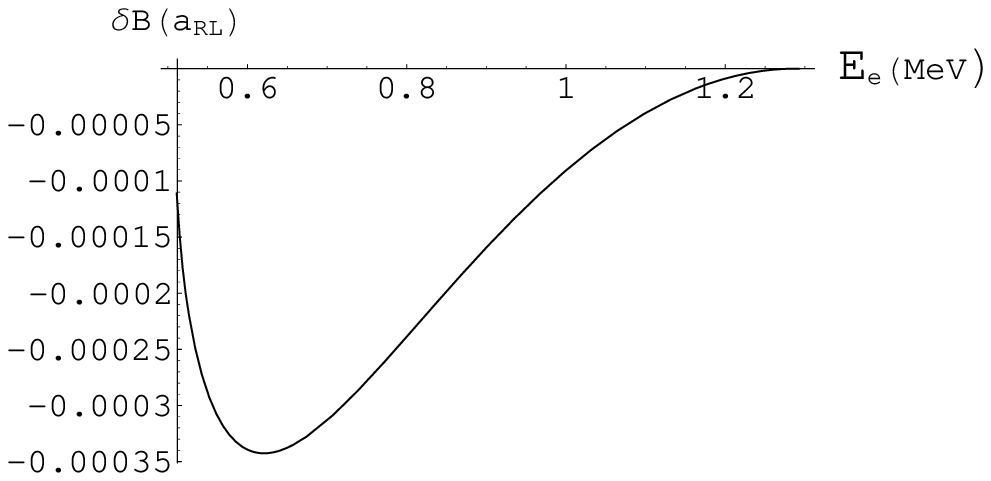}
\caption{Manifestation of $a_{RL}$-type interactions on the $B$ coefficient. }
\label{fig-B-aRL}
\end{figure}

\begin{figure}
\includegraphics{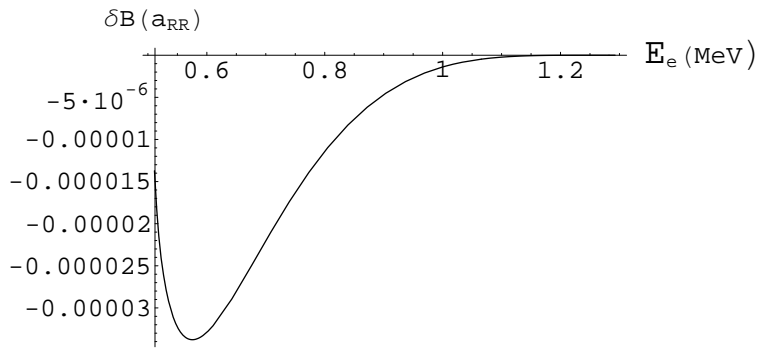}
\caption{Manifestation of $a_{RR}$-type interactions on the $B$ coefficient. }
\label{fig-B-aRR}
\end{figure}

The calculated integral sensitivities of different experiments to a
particular parameter related to new physics can be used for the
estimation of the experimental sensitivity when the experimental
statistics is not good enough. For the optimization of experiments
it is useful to know how manifestations of new physics contribute to the
energy spectrum of the measurable parameter. As an example, the
contributions from $\bar{a}_{LR}$, $\bar{a}_{RL}$ and $\bar{a}_{RR}$
to the spectra for the $a$, $A$ and $B$ correlations are shown on
figures (\ref{fig-a-aLR}) - (\ref{fig-B-aRR}). For uniform presentation all graphs on the figures are normalized by $N_f=G_F^2|V_{ud}|^2 \int f(E) dE$, where $f(E)$ is $a(E_p)$, $A(E_e)$ and $A(E_e)$, correspondingly.
One can see that
these contributions have different shapes and  positions of maxima
both for different model parameters and for different angular
correlations. This gives the opportunity for fine tuning in the
search for particular models beyond the Standard one in neutron
decays.

\begin{figure}
\includegraphics{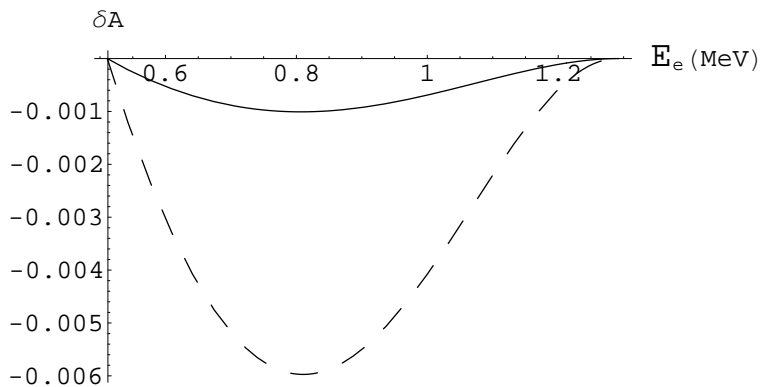}
\caption{Contributions from radiative and recoil corrections (dashed line) and from the left-right model (solid line) to the $A$ coefficient. The curves are explained in the text.}
\label{fig-A-asym}
\end{figure}

\begin{figure}
\includegraphics{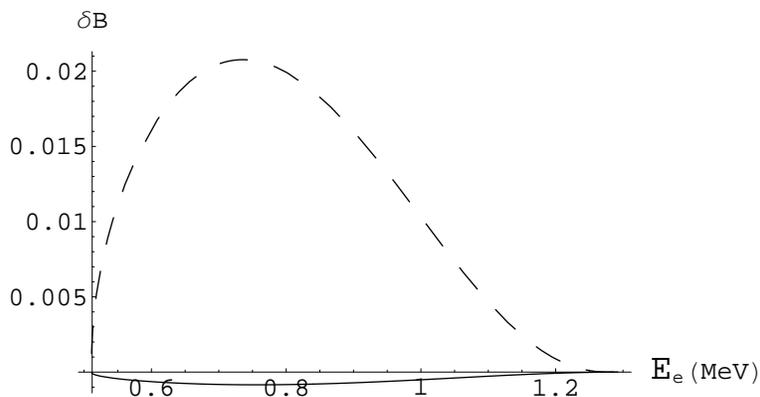}
\caption{Contributions from radiative and recoil corrections (dashed line) and from the left-right model (solid line) to the $B$ coefficient. The curves are explained in the text.}
\label{fig-B-asym}
\end{figure}

Using the  approach developed here one can calculate the exact spectrum for a given model. For example, manifestations of the Left-Right model ($\bar{a}_{RL}= 0.067$ and $\bar{a}_{RR}=0.075$) in the measurements of the $A$ and $B$ coefficients are shown in solid lines on figures \ref{fig-A-asym} and \ref{fig-B-asym}.
 The dashed lines show contributions from recoil effects and radiative corrections (without Coulomb corrections) assuming that $( \alpha /(2 \pi) \; e_V^R) = 0.02$. From these plots one can see the importance of the corrections at the level of the possible manifestations of new physics.
\begin{figure}
\includegraphics{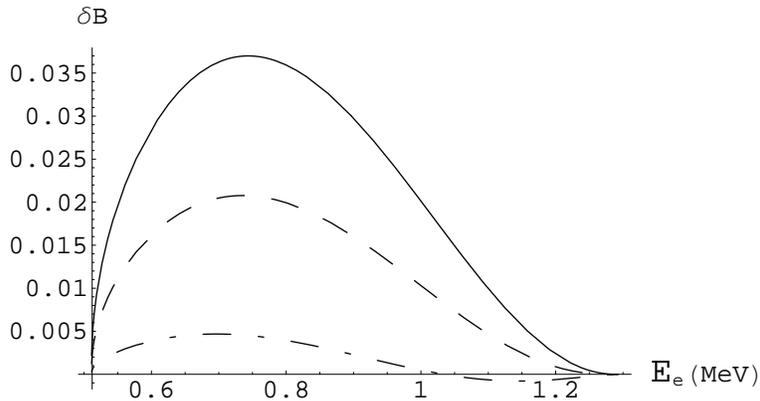}
\caption{Contributions from radiative and recoil corrections to the $B$ coefficient for $ (\alpha /(2 \pi) \; e_V^R)=0.01$ (dashed-doted line), $( \alpha /(2 \pi) \; e_V^R)=0.02$ (dashed line), and $ (\alpha /(2 \pi) \; e_V^R)=0.03$ (solid line).}
\label{fig-B-corr}
\end{figure}
The figure \ref{fig-B-corr} shows how these corrections for the coefficient $B$ affected by the value of the parameter $( \alpha /(2 \pi) \; e_V^R)$ related to nuclear structure: dashed-doted, dashed and solid lines correspond to $0.01$, $0.02$ and $0.03$ values for the parameter.

We presented here results of analysis for only a number of parameters $\bar{a}_{ij}$ to illustrate a different level of sensitivities of experiments to the parameters. For the complete analysis of future experiments all $\bar{a}_{ij}$, $\bar{\alpha}_{ij}$ and $\bar{A}_{ij}$ parameters should be analyzed with a specific experimental conditions taken into account.

\section{Conclusions}

The  analysis presented here provides  a general basis for comparison of different experiments of neutron $\beta$-decay from the point of view of the discovery potential for new physics. It is also demonstrates that various parameters measured in experiments have quite different sensitivities to the detailed nature of the (supposed) new physics and can, in principle be used to differentiate between different extensions to the Standard Model.  Thus neutron decay can be considered as a promising tool to search for  new physics, which may not only detect the manifestations of new physics but also  define the source of the possible deviations from predictions of the Standard model. Our results can be used  for optimization of new high precision experiments to define important directions and to complement high energy experiments. Finally we emphasize that the usual parametrization of experiments in terms of the tree level coefficients $a$, $A$ and $B$, is inadequate when experimental sensitivities are comparable or better to the size of the corrections to the tree level description. This is expected in the next generation of neutron decay experiments. Therefore, such analysis is needed for these experiments. One has to use the full expression for neutron beta-decay in terms of the coupling constants.  In other words, the high precision experiments should  focus on the parameters important for physics rather than on the coefficients $a$, $A$ and $B$ which are sufficient only for low-accuracy measurements.


\begin{acknowledgments}
VG thanks to P. Herczeg for helpful discussions.
This work was supported by the DOE grants no. DE-FG02-03ER46043 and DE-FG02-03ER41258.
\end{acknowledgments}

\end{document}